\documentclass[%
reprint,
superscriptaddress,
amsmath,amssymb,
aps,
prl,
noeprint,nourl,
]{revtex4-2}

\usepackage{graphicx}
\usepackage{dcolumn}
\usepackage{bm}

\usepackage[colorlinks,citecolor=cyan,linkcolor=cyan,urlcolor=cyan,bookmarks=false,hypertexnames=true]{hyperref}

\usepackage[dvipsnames]{xcolor}

\begin{document}
	
	\title{Deterministic Vortex Generation from Coupled Trapped Polariton Condensate Triad}
	
	\author{X. Qin}
	\email{qinxiao@westlake.edu.cn}
	\affiliation{Department of Physics, Fudan University, Shanghai 200433, China}
	\affiliation{Department of Physics, School of Science, Westlake University, Hangzhou 310014, China}
	
	\author{N. Pitanios}
	\affiliation{IESL and Center for Quantum Science and Technologies, FORTH, 70013 Heraklion, Greece}
	\affiliation{Department of Materials Science and Engineering, University of Crete, Heraklion, Greece}
	\affiliation{Department of Physics, School of Science, Westlake University, Hangzhou 310014, China}
	
	\author{S. Dam}
	\affiliation{Julius-Maximilians-Universit\"{a}t W\"{u}rzburg, Physikalisches Institut and W\"{u}rzburg-Dresden Cluster of Excellence ctd.qmat, Am Hubland, W\"{u}rzburg, Bavaria, Germany}
	
	\author{S. Betzold}
	\affiliation{Julius-Maximilians-Universit\"{a}t W\"{u}rzburg, Physikalisches Institut and W\"{u}rzburg-Dresden Cluster of Excellence ctd.qmat, Am Hubland, W\"{u}rzburg, Bavaria, Germany}
	
	\author{S. H\"{o}fling}
	\affiliation{Julius-Maximilians-Universit\"{a}t W\"{u}rzburg, Physikalisches Institut and W\"{u}rzburg-Dresden Cluster of Excellence ctd.qmat, Am Hubland, W\"{u}rzburg, Bavaria, Germany}
	
	\author{S.V. Koniakhin}
	\email{kon@rqc.ru}
	\affiliation{Russian Quantum Center, Skolkovo, Moscow 121205, Russian Federation}
	
	\author{P.G. Savvidis}
	\affiliation{Department of Physics, School of Science, Westlake University, Hangzhou 310014, China}
	\affiliation{IESL and Center for Quantum Science and Technologies, FORTH, 70013 Heraklion, Greece}

	\date{\today}
	
	\begin{abstract}
		
		We present the deterministic generation and all-optical switching of ternary topologically distinct vortex states in a triad of trapped exciton-polariton condensates with direct state readout from momentum-space emission distribution. Tuning the non-resonant pump power controls the sign of the dissipative coupling, selecting between a ferromagnetic in-phase state and two frustrated states with global phase winding. The zero-winding ferromagnetic state has unstructured far-field emission at low momenta and forms vortex-antivortex arrays at the Dirac points, reflecting the $C_{3\mathrm{v}}$ global symmetry. The two frustrated states of $\pm 2\pi$ winding can be activated on-demand by a control beam and produce different triangular emission patterns selectively localized at the $K$ or $K'$ points. Although all three states share identical real-space emission distributions, their momentum-space photoluminescence profiles distinguish them unambiguously, revealing the chirality without the need for interferometry. This enables on-demand orbital angular momentum control using a pump independent of resonant phase or OAM imprinting, establishing condensate triads as a compact platform for reconfigurable chiral light generation operating on a ternary logic basis.

		
	\end{abstract}
	
	\maketitle

	\textit{Introduction.}---Vortices in nonequilibrium exciton-polariton systems~\cite{lagoudakis2008quantized,roumpos2011single}--quantum fluids of light\cite{Carusotto2013, bloch2022non}--share the defining features of rotation quantum topological excitations that permeate diverse branches of physics, from superfluids and superconductors to nonlinear optics \cite{onsager1949statistical,abrikosov1957magnetic} and particle physics~\cite{maksimov2025diffraction}. Characterized by quantized phase winding around a density-depleted core, these excitations carry orbital angular momentum (OAM) and govern phenomena from quantum turbulence \cite{bradley2012energy,Gauthier2019,Johnstone2019,panico2023onset} to Berezinskii-Kosterlitz-Thouless phase transitions \cite{roumpos2012power,Caputo2018}. They arise as stable solutions of both conservative and driven-dissipative Gross-Pitaevskii equations~\cite{lagoudakis2008quantized,sigurdsson2017parity,claude2020taming} and can appear spontaneously due to intrinsic system defects ~\cite{lagoudakis2008quantized} or symmetry breaking~\cite{ohadi2016nontrivial}. Considerable effort has been devoted to their control as possible units for optical information encoding and manipulation. Significant progress has been achieved in circular trap geometries~\cite{dall2014creation}, where on-demand switching between opposite winding signs has been realized~\cite{ma2020realization}, and in the imprinting of OAM via static Laguerre-Gauss~\cite{kwon2019direct} and rotating potential~\cite{del2023optically,gnusov2023quantum} excitation.

	Condensates arranged in polygon geometries have also emerged as a versatile platform for studying spontaneous vortex formation through geometric frustration~\cite{cookson2021geometric}. For odd-numbered polygons with antiferromagnetic coupling, geometric frustration prevents simultaneous anti-phase locking of all nearest neighbors, leading to spontaneous symmetry breaking and circulating polariton currents. Yet in the elementary building block of such arrays, the condensate triad, vortex-carrying configurations were absent in Refs.~\cite{cookson2021geometric,tosi2012geometrically}, while signatures of vortex states reported in Ref.~\cite{ohadi2016nontrivial} appeared only as a statistical mixture of the two opposite winding directions. Deterministic selection of the winding direction, and switching between topologically distinct states of the triad, has not been demonstrated.
	
	
	The reciprocal (momentum) space provides an equally rich arena for vortex physics, where topological configurations can arise from band structure singularities~\cite{bena2009remarks,fuchs2010topological,lim2015geometry}. Recently, an alternative route to momentum-space vortex generation was demonstrated where excitation of an optical honeycomb lattice in atomic vapor cell with a simple unstructured Gaussian beam produced regular arrays of vortex-antivortex pairs near the Dirac points in momentum-space~\cite{li2023simultaneous}. The underlying mechanism is remarkably general. The $C_{3\mathrm{v}}$ symmetry of a wave packet localized on a honeycomb site with three nearest neighbors generates momentum-space phase winding directly through the Fourier transform, analogous to the phase factors of the tight-binding description of graphene~\cite{bena2009remarks}, and requires no Berry curvature or lattice topology.

	Here, we bridge these two approaches. We realize a triad of exciton-polariton condensates in optically imprinted hexagonal traps, which offer a lower condensation threshold and extended coherence compared to ballistic condensates ~\cite{cristofolini2013optical,brune2025quantum}, while preserving effective coupling between distant sites \cite{ohadi2016tunable} unlike the circular trapping potential~\cite{ma2020realization}, which would be essential for implementation of polariton simulators. The triad constitutes the minimal $C_{3\mathrm{v}}$ symmetric unit of the honeycomb/hexagonal lattice, so the symmetry driven mechanism of momentum-space vortex generation demonstrated in passive photonic lattices~\cite{li2023simultaneous} applies directly. Crucially, unlike passive lattices, the polariton system is non-Hermitian, and the sign of the dissipative coupling between condensates is controlled all-optically by the pump power, allowing deterministic selection among $0$ and $\pm 2\pi$ global phase winding.
	
	We switch on demand between the in-phase configuration, producing vortex-antivortex arrays at the Dirac points~\cite{li2023simultaneous}, and the frustrated configuration with $\frac{2\pi}{3}$ phase differences between adjacent condensates \cite{cookson2021geometric}, whose global chirality is set by the position of a control beam. This replaces the stochastic chirality selection of Ref.~\cite{ohadi2016nontrivial}. Owing to the particular trapped geometry and size of the system, the orbital angular momentum and vortex state are directly mapped onto the momentum-space emission intensity, allowing them to be read out without interferometry or an OAM sorting scheme, opening pathways for polariton OAM devices, vortex-mediated information encoding in ternary basis, and controlled studies of vortex dynamics in quantum fluids of light.

	\begin{figure}
		\centerline{\includegraphics[width=0.95\columnwidth]{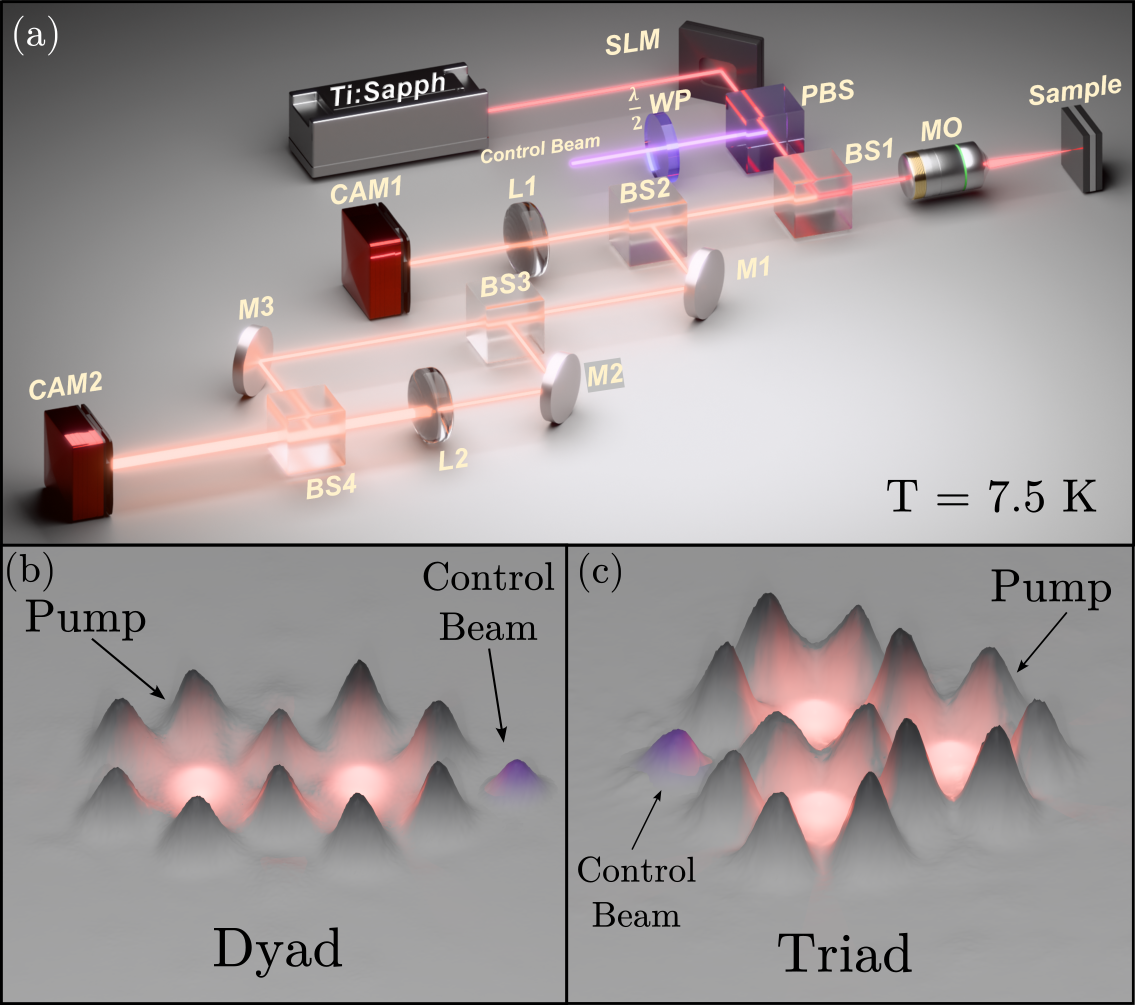}}
		
		\caption{(a) Schematic depiction of the experimental setup for generating and detecting optically trapped polariton condensates. (b,c) Spatially structured excitation patterns for different condensate arrangements: (b) two-condensate configuration and (c) three-condensate configuration. The purple spot indicates the control beam.}
		\label{fig:figure_1}
	\end{figure}

	
	
	
	
	
	\textit{Experimental.}--The experiments were performed on a high-quality AlAs/Al$_{x}$Ga$_{(1-x)}$As $(x = 0.15)$ planar semiconductor microcavity\cite{savvidis2000angle,kavokin2017microcavities} with four sets of three GaAs quantum wells, held at 7.5~K. The sample is in the strong light-matter coupling regime, at an exciton-photon detuning of $\delta = -7.5$~meV. Polariton condensates were excited non-resonantly with a single-mode continuous-wave Ti:sapphire laser at 750~nm, tuned to the first Bragg mode 
	of the mirror stop-band.

	As shown in Fig.~\ref{fig:figure_1}(a), the excitation beam is shaped by a spatial light modulator (SLM) and projected onto the sample through a microscope objective, while the polariton photoluminescence is routed to the imaging and interferometric detection arms. The SLM generates the optically imprinted excitation landscape defining the trapping potential. The corresponding geometries are illustrated in Figs.~\ref{fig:figure_1}(b,c), showing the two-condensate dyad and three-condensate triad configurations, respectively. In both cases, the condensates are confined within an optically imprinted hexagonal lattice potential, while an additional localized control beam [purple in Figs.~\ref{fig:figure_1}(b,c)] provides independent optical tuning of the reservoir. The SLM hologram was iteratively optimized using a conjugate-gradient algorithm combined with real-time camera feedback~\cite{schroff2023accurate}, compensating for the Gaussian excitation profile and ensuring high intensity homogeneity across the lattice sites.
	
	Achieving phase locking between trapped polariton condensates is challenging due to the stochastic nature of spontaneous condensation, pump noise, and sample inhomogeneities. Local optical modification of the exciton reservoir has previously been employed to control the coupling between spatially separated polariton condensates~\cite{PhysRevLett.124.207402}. Here, a secondary control beam, cross-polarized with respect to the primary non-resonant pump, is used to extend this concept to condensates confined within an optically imprinted potential. Since the two beams do not interfere, the control beam locally reshapes the exciton reservoir without perturbing the main trap, providing fine all-optical control of the potential landscape and the gain profile, and thereby enabling robust phase locking in both the condensate dyad and triad.
	
	The phase configurations of the polariton dyads and triads were characterized using a modified Mach-Zehnder interferometer, following the interferometric phase-retrieval approach employed in Ref.~\cite{doi:10.1126/sciadv.ado4042}. Here, we extend this technique to reconstruct the condensate phase in both real- and momentum ($k$) space. In real space, the condensate emission was interfered with a magnified, nearly uniform reference obtained by expanding the emission from one of the condensates, while in momentum-space the reference was generated by expanding the brightest lobe of the emission pattern. Momentum-space intensity distributions, monitored in parallel on a separate camera, provided unambiguous rapid identification of the target configurations, which were subsequently confirmed interferometrically. The interferograms were processed with an off-axis Fourier-transform algorithm~\cite{Takeda:82}, yielding full two-dimensional intensity-weighted phase maps and canonical determination of the phase relationships among the condensates.
	
	
	
	\textit{Theoretical background.}---The single polariton condensate can be characterized with a scalar complex-valued amplitude $A$ that obeys the Stuart-Landau equation $\frac{\partial}{\partial t}A = \lambda(P) A - c |A|^2A$, where the real part of $\lambda(P)$ is the growth rate of the condensate during condensation at a certain power $P$ and the imaginary part is the energy of the condensate~\cite{utesov2025universal, utesov2026subcritical}. The complex parameter $c$ describes reservoir-mediated non-linear effects including depletion at high condensate densities (real part) and energy blue-shift (imaginary part). In case of neighboring condensates, the coupling $T_{ij} = iJ_{ij} + D_{ij}$ composed of tunneling (Josephson) and dissipative parts can be naturally introduced to 
	\begin{subequations}
		\begin{align}
			\frac{dA_1}{dt} &= \lambda_1' A_1 + T_{12}A_2 + T_{13}A_3 - c_1|A_1|^2A_1 \\
			\frac{dA_2}{dt} &= \lambda_2' A_2 + T_{21}A_1 + T_{23}A_3 - c_2|A_2|^2A_2 \\
			\frac{dA_3}{dt} &= \lambda_3' A_3 + T_{31}A_1 + T_{32}A_2 - c_3|A_3|^2A_3
			\label{eq_stl}
		\end{align}
	\end{subequations}
	
	From the acquired amplitudes $A_i$, one can reconstruct the full polariton field (order parameter) profile as $E(\mathbf{r}) = \sum_i A_i \psi_i(\mathbf{r})$, where $\psi_i(\mathbf{r})$ is the Gaussian profile of a single condensate. The final states depend on the initial random seed with statistical prevalence of certain ones depending on particular parameters.
	
	The expected to observe in experiment states are evident via the hopping matrix:
	
	\begin{equation}
		\label{eq:linear}
		\mathbf{T}
		= (D + iJ)
		\begin{pmatrix}
			0 & 1 & 1 \\
			1 & 0 & 1 \\
			1 & 1 & 0
		\end{pmatrix}
	\end{equation}
	
	The eigenmode with larger growth rate would dominate in the experimental photoluminescence pattern~\cite{cao2025polariton}. So, the ferromagnetic all-in-phase mode $(1,1,1)^T$ corresponding to eigenvalue $2(D + iJ)$ is a signature positive dissipative coupling $D$. The two degenerate modes $(1, e^{\pm i\frac{2\pi}{3}}, e^{\pm i\frac{4\pi}{3}})$ with eigenvalue $-(D + iJ)$, which represent vortex- and anti-vortex-like phase singularities~\cite{cookson2021geometric}, are favorable for negative dissipative coupling. The non-linear effects above the threshold lead to selection between the opposite-chirality modes, which is captured by randomly-seeded Stuart-Landau model. In experiment, pumping power and distances define the actual condensate couplings. In the closely related system of the two spatially separated ballistic condensates, polariton dyad~\cite{topfer2020time}, either ferromagnetic configuration is observed at positive dissipative coupling (mode $(1,1)^T$ and eigenvalue $D + iJ$) or the antiferromagnetic one at negative dissipative coupling (mode $(1,-1)^T$ and eigenvalue $-D - iJ$).

	

	
	\begin{figure}
		\centering	
		\includegraphics[width=\linewidth]{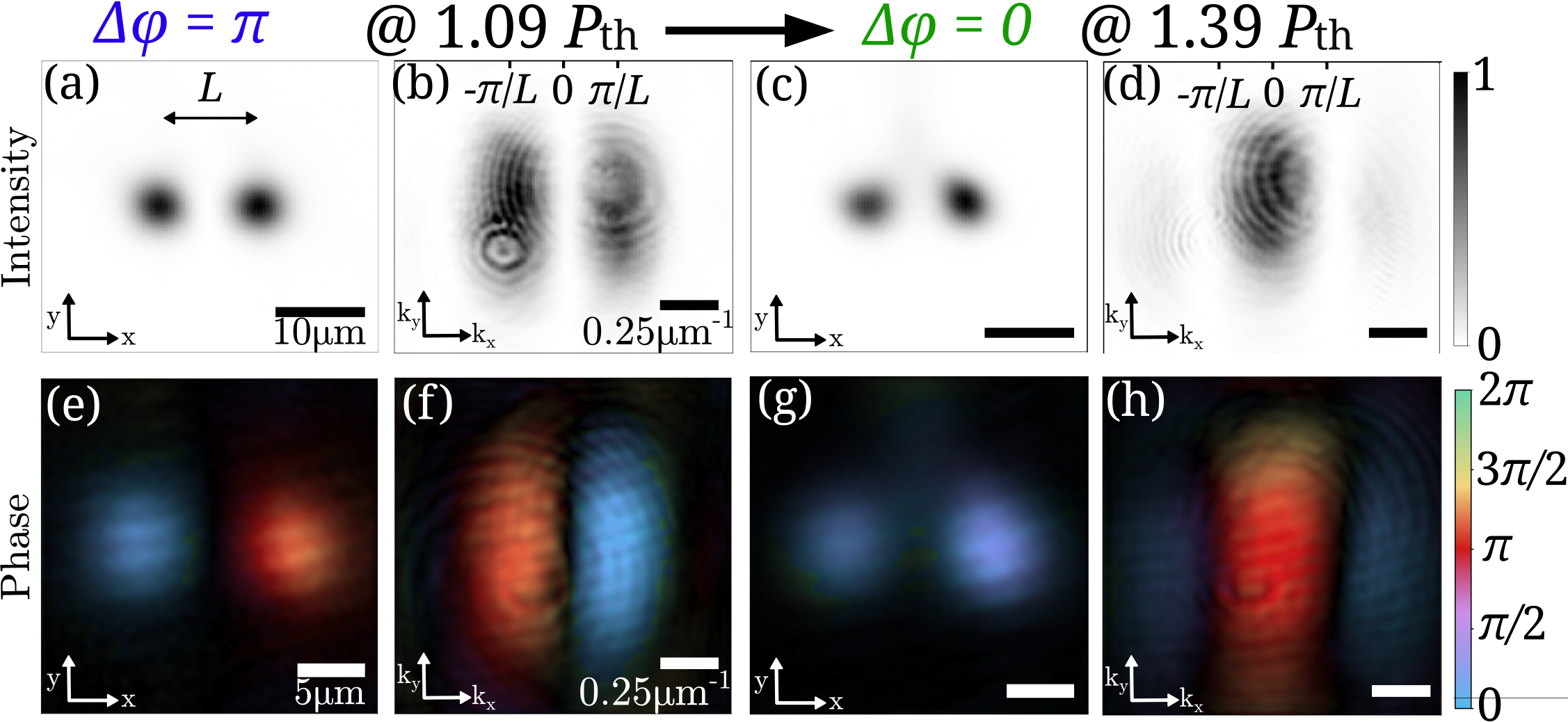}
		\caption{Polariton dyad in anti-symmetric and symmetric configurations. Panels (a) and (c) map the real-space PL intensity for the anti-symmetric (negative dissipative coupling) and symmetric (positive dissipative coupling) states, while panels (b) and (d) show the respective momentum-space intensities. The lower panels display the corresponding phase profiles: panels (e) and (g) capture the real-space phase, and panels (f) and (h) depict the momentum-space phase distributions. The combination of zero intensity and $\pi$ phase jump in momentum-space resemble the characteristic soliton patterns.}
		\label{fig:figure_2}
	\end{figure}

	
	
	
	
	\textit{Polariton dyad.}---We first characterize the elementary building block, a polariton condensate dyad, in which the pair of condensates~\cite{topfer2020time} is synchronized via fine tuning of the pump power of the incoherent optical traps. The pump power precisely modulates the dissipative coupling $D$, driving a transition from an antiferromagnetic (out-of-phase, $\pi$) state slightly above threshold to a ferromagnetic (in-phase, 0) state at higher power. Figure~\ref{fig:figure_2} shows the real- and momentum-space intensity and phase distributions of the dyad at the two pump powers. For the antiferromagnetic state at $1.09~P_{\rm th}$, the momentum-space pattern [Fig.~\ref{fig:figure_2}(b)] displays zero intensity at the center, reflecting the $\pi$ phase difference between the condensates. The wave function, having opposite signs at the two sites, has zero overlap with low-momenta Fourier harmonics and reaches its maximum at the edges of the generated Brillouin zone, located at $\pm\frac{\pi}{L}$, where $L$ denotes the spacing between the condensates. The extracted phase maps [Figs.~\ref{fig:figure_2}(e,f)] confirm this directly, revealing the $\pi$ phase difference between the condensates in real space and between the lobes in momentum-space.
	
	Increasing the pump power to $1.39~P_{\rm th}$ reverses the sign of the effective coupling, and the dyad switches to in-phase locking, unambiguously recognized by the lobe at $k=0$ [Fig.~\ref{fig:figure_2}(d)]. Notably, the phase of the momentum-space PL in the nearest Brillouin zone replicas is shifted by $\pi$ with respect to the central lobe [Fig.~\ref{fig:figure_2}(h)]. Zero density accompanied by an abrupt $\pi$ phase jump is the well known fingerprint of the 1D soliton, the simplest topological defect existing in effectively one-dimensional systems. Because full information on the crystal structure and its reciprocal lattice is already encoded in the dyad, referring to the concept of Brillouin zone and its high symmetry points is practical. The same principles of Fourier transformation~\cite{boguslawski2011increasing,li2023simultaneous} govern the phase patterns over the Brillouin zone replicas of the condensate triad, where the 2D geometry admits more sophisticated topological defects, namely vortices.
	
	
	
	\textit{Polariton triad.}---This all-optical inversion of the coupling from antiferromagnetic to ferromagnetic provides the control we now exploit in the triad. Extending it to a triangular condensate array, trapped within the three-hexagon lattice, provides the minimal geometric system for exploring polariton networks with topologically distinct states and nontrivial OAM of the emitted light. For a phase-locked polariton triad, we observe the realization of the essential phase configurations encoded in the modes of the hopping matrix in Eq.~\eqref{eq:linear}, an all-in-phase ferromagnetic state and frustrated chiral states characterized by global $\pm 2\pi$ winding in real and momentum-space.

\begin{figure}
		
		\centerline{\includegraphics[width=\columnwidth]{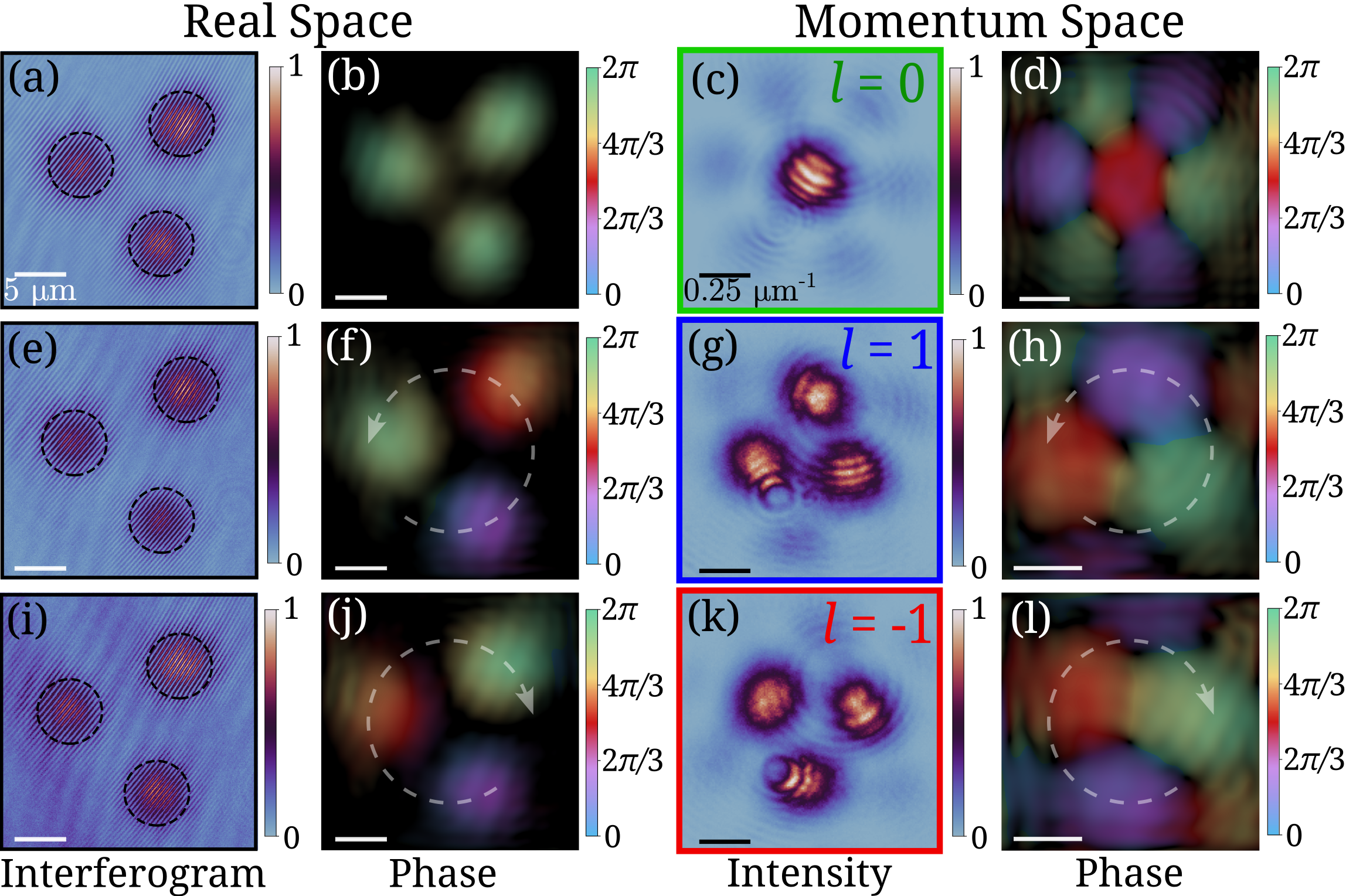}}
		\caption{Real and momentum-space intensity and retrieved phase for polariton triad from the experiment. Panels in the upper row (a)-(d) are for the all-in-phase configuration with zero global winding. Panels (e)-(h) and (i-l) are for opposite nonzero winding direction of global phase in frustrated configuration. Phase difference between condensates in real space (f),(j) and lobes in momentum-space (d),(h),(l) is approx. $\pm\frac{2\pi}{3}$. Despite similar PL profile in real space, all winding states $l = -1,0,1$ can be unambiguously distinguished by momentum-space (far field) intensity distribution shown in panels (c),(g),(k).}
		\label{fig:tri_exp}
	\end{figure}

	In the all-in-phase structure characterized by positive dissipative coupling, the real space interferograms exhibit zero phase shift across the lattice sites, shown in Fig.~\ref{fig:tri_exp}. In momentum-space, PL emission intensity is mainly concentrated in the first Brillouin zone because the signal from three condensates is in superposition for the low-momenta Fourier harmonics. The additional regular hexagon of weaker emission spots from BZ replicas is clearly visible. The momentum-space phase pattern, is richer, i.e. the phase is homogeneous over the Brillouin zone, $K$ Dirac points carry vortices and their $K'$ counterparts carry anti-vortices~\cite{li2023simultaneous}. In accordance with the case of dimer, ferromagnetic configuration is characteristic for higher pump intensity.

	The reciprocal lattice and BZ is here also essentially generated by a triad, being the simplest $C_{3\mathrm{v}}$-symmetric wave function on the real-space hexagonal lattice. Momentum-space PL signal that can be detected from BZ replicas has a cutoff $\approx \frac{\pi}{D}$, because the Fourier integrals vanish for plane waves shorter than the condensate size $D$. The size of BZ itself estimates via the condensate spacing as $\frac{\pi}{L}$. In the present experiment, few BZ replicas are visible because the condensate spacing is approximately three times larger than their size.
	

	At lower pumping power, the dissipative coupling is negative, leading to frustration because integer phase jumps for all three condensates are not allowed. As a result, one sees fractional, approx. $\pm\frac{2\pi}{3}$, phase jumps between the condensates in real space. The global phase winding around the center is either $2\pi$ or $-2\pi$. Importantly, the two opposite winding are indistinguishable in real space emission profile, as well as for the all-in-phase configuration, but for momentum-space they are characterized by mirror reflected patterns~\cite{maksimov2025diffraction}, i.e. triangular-shape profiles with maxima either in $K$ or $K'$ points. The full phase circulation in momentum-space is the same as in real space. In the experiment, one can switch on-demand between opposite winding by tuning the control beam position, which breaks the chiral symmetry and releases the degeneracy in mode growth rate.
	
	The actual size and spacing of the condensates allow fine resolution of momentum-space and straightforward interpretability while ballistic condensate chains produce sophisticated structures~\cite{cookson2021geometric}. As a result, one can directly and unambiguously judge about the global phase in condensate triad from momentum-space PL intensity without additional interferometry or OAM sorting.
	
\begin{figure}
		
		\centerline{\includegraphics[width=\columnwidth]{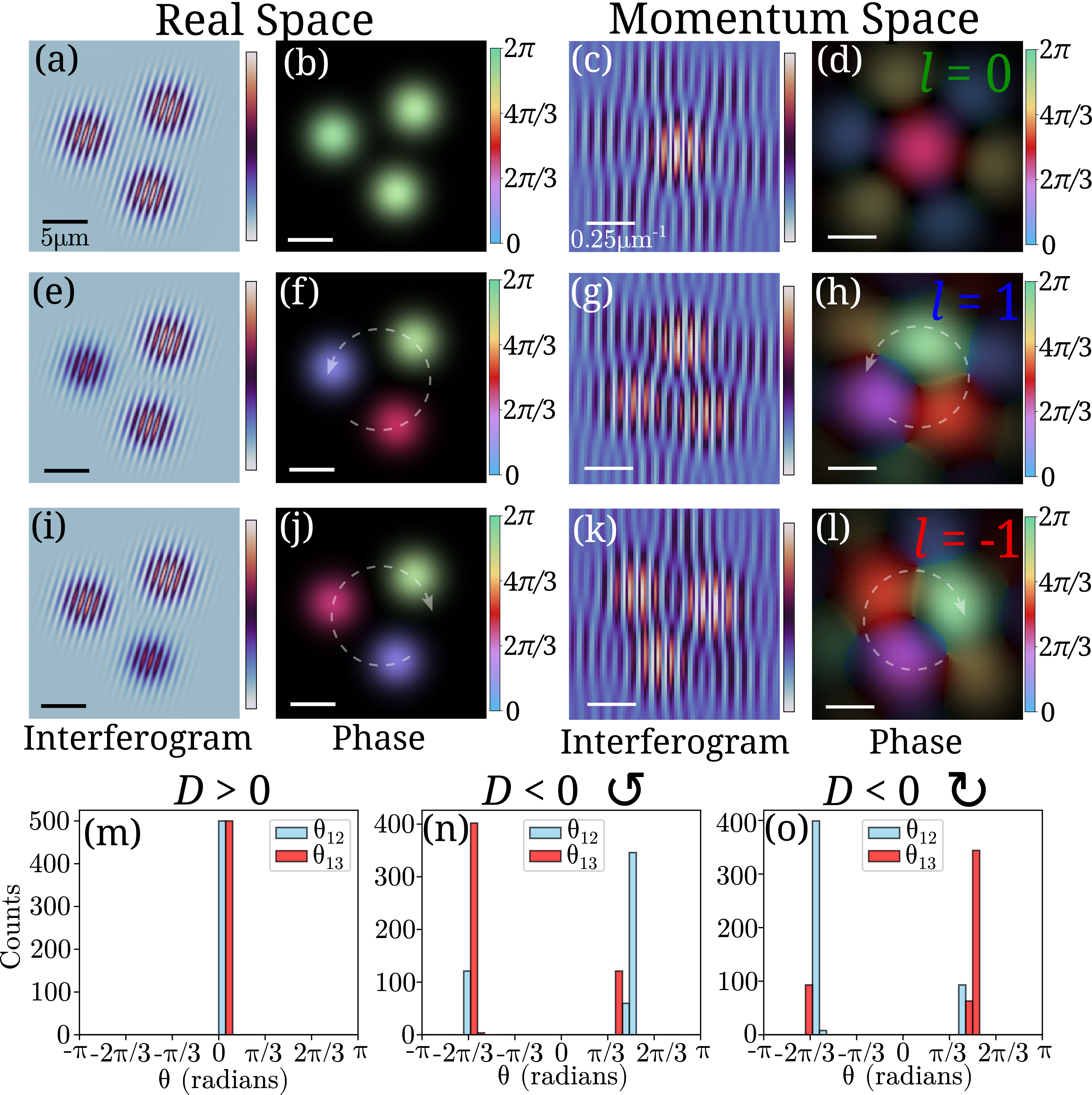}}
		
		\caption{Results of Stuart-Landau equation simulations reproducing Fig.~\ref{fig:tri_exp}. The growth rate $\Re(\lambda)$ and saturation constant $\Re(c)$ responsible for the single condensate amplitudes are 0.75 and 0.9, respectively. $\Im(c) = 0.5$. Coupling is given by $|J| = |D| = 0.15$. For ferromagnetic configuration in panels (a)-(d), dissipative coupling $D>0$. The opposite vortices in the antiferromagnetic configuration at $D<0$ are characterized by clockwise (e)-(h) and anticlockwise (i)-(l) ordering of site energies: $\Im({\lambda_1,\lambda_2,\lambda_3}) = (0.0,0.2,0.14)$ and $\Im({\lambda_1,\lambda_2,\lambda_3}) = (0.0,0.14,0.2)$, respectively. Panels (m)–(o) show final phase histograms for condensates 2 and 3 with respect to condensate 1 across the three parameter sets.}
		\label{fig:tri_stl}
	\end{figure}

	The transition between the ferromagnetic and frustrated configurations can be fully reproduced within the coupled Stuart--Landau oscillator model by reversing the sign of dissipative coupling $D$ (see Fig.~\ref{fig:tri_stl}). The condensate order parameter $E(\bf r)$ was reconstructed for multiple random initial seeds of $A_{1,2,3}$ and subsequently analyzed following the experimental procedure of phase extraction. When the dissipative coupling is positive, the all-in-phase configuration emerges as the sole final state, independent of the initial conditions. For negative coupling, robust observation of a configuration with a well-determined winding sign requires degeneracy lifting between opposite winding and thus a chirality source in the system must be introduced. It was found that gradual clockwise or anticlockwise increase in the site energies $\Im(\lambda_{1,2,3})$ was the most effective way. In the absence of chirality, the system becomes bistable with the winding direction of the final state defined by the stochastic initial seed, thus yielding a statistically balanced mixture of left- and right-winding final states rather than a coexistence of left and right vortices in the limit-cycle regime. Key system features---including the all-in-phase configuration at higher pump power, as well as bistability, the absence of limit cycles, and chiral symmetry breaking at lower power---are all reproduced within the mean-field dynamics (see the Supplemental Material~\cite{SM}).

	\textit{Conclusions.}---We have demonstrated deterministic all-optical control of the real- and momentum-space OAM of light emitted by a triad of trapped exciton-polariton condensates. Both the onset of phase locking and the precise phase ordering are directly visible in the momentum-space emission, despite identical real-space intensity profiles. The ferromagnetic in-phase configuration focuses its far-field emission into the first induced Brillouin zone, featuring vortex–antivortex pairs at the Dirac points. In contrast, the two frustrated configurations, which possess a global winding of $\pm2\pi$, give rise to mirror-image triangular patterns at the $K$ and $K'$ points, respectively. Consequently, this figure enables the winding sense to be mapped directly onto a ternary $l = -1, 0 ,+1$ basis of three distinct states using only the momentum-space PL intensity distribution, eliminating the need for interferometry~\cite{dall2014creation} or orbital angular momentum sorting~\cite{ma2020realization}. The synchronization is governed by the sign of the non-Hermitian dissipative coupling, set by the pump intensity, with positive coupling selecting the in-phase state and negative coupling the winding-carrying states, while the cross-polarized control beam compensates sample inhomogeneities and deterministically selects between the two winding directions. In contrast to the ballistic regime, where superfluid flows can strongly deform the condensates~\cite{ohadi2016nontrivial, cookson2021geometric}, and to isolating circular traps that would attenuate the adjacent vortex cross-talk, the trapped geometry preserves high-quality Gaussian modes, making the triad, the minimal element encoding the full symmetry of the hexagonal lattice, a compact versatile source of reconfigurable OAM light switchable by excitation power alone.
	
	\begin{acknowledgments}
		\textit{Acknowledgments.}---P.G.S. acknowledges support from the Innovation Program for Quantum Science and Technology (2023ZD0300300), the Innovation Resource Allocation of Westlake University-Zhejiang Provincial Natural Science Foundation of China under Grant No. XHD24A2401, and the Westlake University project No. 102510036022301. S.D., S.B. and S.H. acknowledge support by the State of Bavaria and the German Research Foundation (DFG) within the project "OAM-SLED" (project id HO 5194/20-1).\\
		
	\end{acknowledgments}

	
	\bibliography{bib}

	\onecolumngrid
	\appendix

\clearpage

\setcounter{figure}{0}
\renewcommand{\thefigure}{S\arabic{figure}}

\section{Equiprobabilistic bistable configuration}

\begin{figure}[h]
	\centerline{\includegraphics[width=\columnwidth]{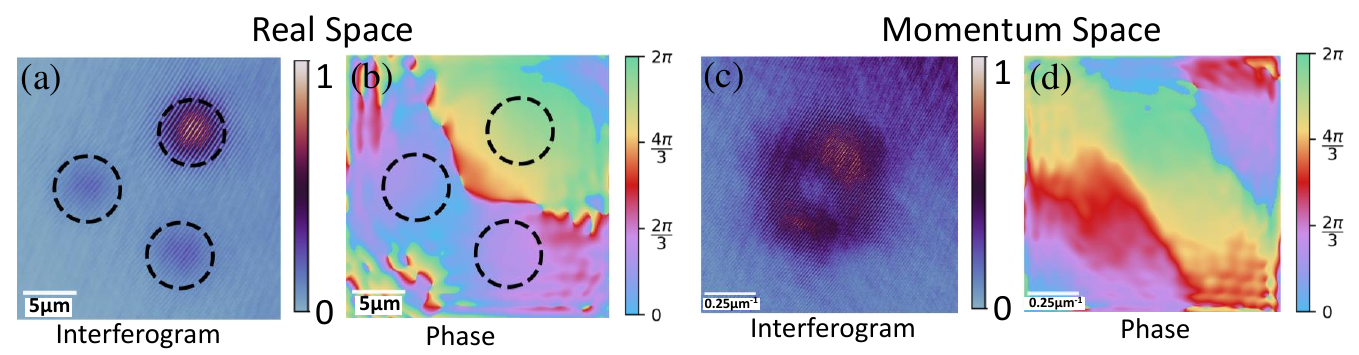}}
	\caption{Experimental observation of bistability with equiprobable opposite winding directions.}
	\label{fig:S1}
\end{figure}

\begin{figure}[h]
	\centerline{\includegraphics[width=\columnwidth]{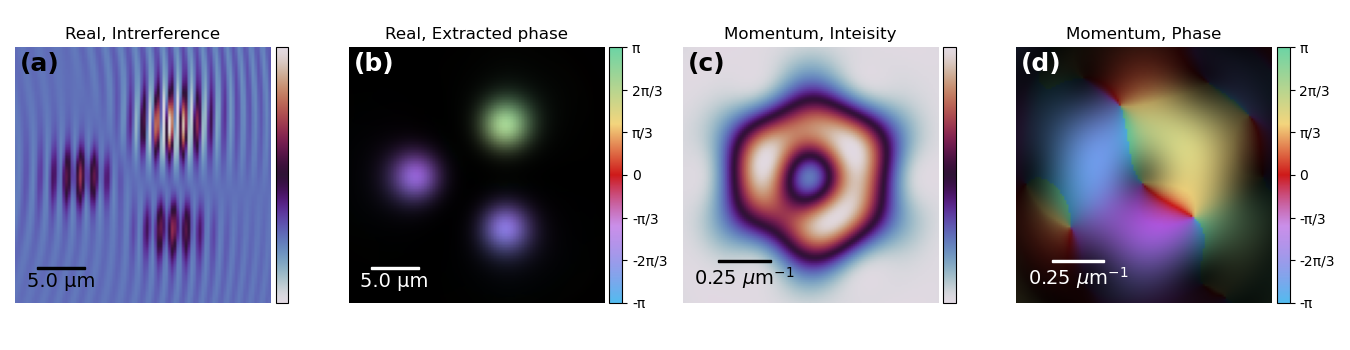}}
	\caption{Bistable configuration within Gross-Pitaevskii approach with equivalent probability of opposite winding directions in frustrated configuration. The images are the result of averaging over many runs with initial noise-like seed with the phase retrieval fully reproducing the experiment.}
	\label{fig:S2}
\end{figure}

\begin{figure}[h]
	\centerline{\includegraphics[width=0.99\columnwidth]{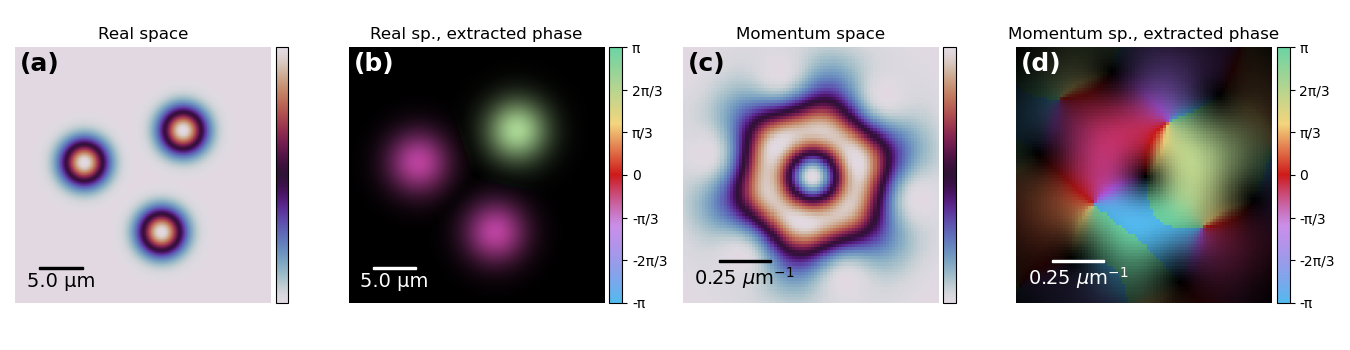}}
	\caption{Bistable configuration within Stuart-Landau equation approach with equivalent probability of opposite winding directions in frustrated configuration.}
	\label{fig:S3}
\end{figure}

To distinguish the bistable vortex state from the deterministic vortex state, Figs.~\ref{fig:S1}--\ref{fig:S3} summarize their characteristic differences. Specifically, Fig.~\ref{fig:S1} presents the experimental observation of the bistable state, which is captured by mean-field ddGPE numerical simulations (Fig.~\ref{fig:S2}) and theoretical calculations based on the Stuart-Landau equation (Fig.~\ref{fig:S3}). 

Momentum-space intensity distribution serves as a crucial signature to discriminate between these two regimes: as shown in Figs.~\ref{fig:S1}(c)--\ref{fig:S3}(c), the bistable state manifests as an O-ring-like pattern in momentum space, whereas the deterministic state yields an unambiguous vortex signature of tringular emission in $K$ or $K'$ points as demonstrated in the main text.

\clearpage

\section{Experimentally Selecting Vortex direction}

In the experiment, deterministic chirality is introduced into three coupled condensates via the  to explicitly break the system symmetry. This leads to the emergence of a well-defined phase chirality, where the direction of the phase vortex aligns with the intensity winding direction.

\begin{figure}[h]
	\centerline{\includegraphics[width=\columnwidth]{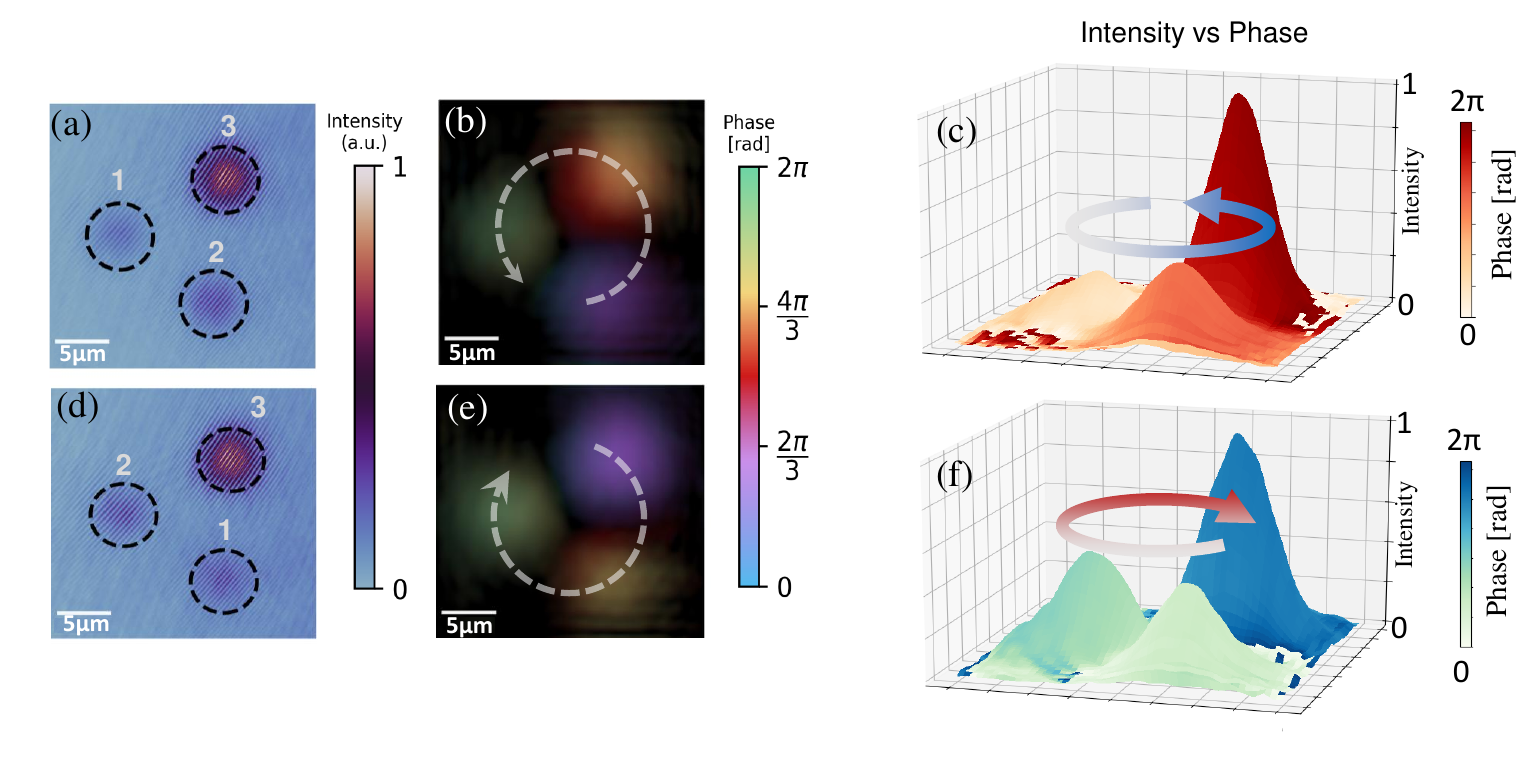}}
	\caption{Experimental demonstration of co-directional and counter-directional intensity-phase chirality. (a) Measured interferogram of three coupled condensates, with spots labeled 1, 2, and 3 in increasing order of light intensity. (b) Spatial phase map extracted from the interferogram in (a).Panel (c) shows 3D surface plot displaying both intensity profile and extracted phase (shifted vertically by an offset for visual clarity), demonstrating that the phase gradient aligns with the intensity winding direction. (d–f) Corresponding set of interferogram, phase map, and 3D profile for a state with opposite chirality, showing counter-directional alignment.}
	\label{fig:S4}
\end{figure}

\clearpage

\section{Gross-Pitaevskii equations for exciton-polairotns}

The system dynamics is governed by a driven-disipative Gross-Pitaevskii (mean-field) equation for the polariton order parameter $\psi(\mathbf{r}, t)$ coupled to a rate equation for the exciton reservoir density $n_R(\mathbf{r}, t)$.

The evolution of the polariton field is given by
\begin{equation}
i\hbar\frac{\partial \psi(\mathbf{r}, t)}{\partial t} = \left[ -\frac{\hbar^2}{2m_{\text{pol}}}\nabla^2 + V(\mathbf{r}, t) - \frac{i\hbar}{2\tau_{\text{cav}}} \right] \psi(\mathbf{r}, t) + i \frac{\hbar R}{2} n_R(\mathbf{r}, t) \psi(\mathbf{r}, t) + i\hbar \sigma \xi(\mathbf{r}, t),
\label{eq:polariton}
\end{equation}
where the effective potential $V(\mathbf{r}, t)$ includes polariton-polariton and polariton-reservoir interactions:
\begin{equation}
V(\mathbf{r}, t) = \alpha_1 |\psi(\mathbf{r}, t)|^2 + \alpha_{1R} \left( n_R(\mathbf{r}, t) + \eta P(\mathbf{r}) \right).
\label{eq:potential}
\end{equation}
Here, $\xi(\mathbf{r}, t)$ is a complex Gaussian white noise term with zero mean and correlation $\langle \xi(\mathbf{r}, t) \xi^*(\mathbf{r}', t') \rangle = 2\delta(\mathbf{r}-\mathbf{r}')\delta(t-t')$, representing quantum fluctuations. The dimensionless coefficient $\sigma$ controls the noise amplitude.

The exciton reservoir density obeys the rate equation
\begin{equation}
\frac{\partial n_R(\mathbf{r}, t)}{\partial t} = P(\mathbf{r}) - \left( \frac{1}{\tau_R} + \frac{R}{\hbar} |\psi(\mathbf{r}, t)|^2 \right) n_R(\mathbf{r}, t),
\label{eq:reservoir}
\end{equation}
where $P(\mathbf{r})$ is the spatially dependent pump profile.

The temporal integration of Eqs.~\eqref{eq:polariton} and \eqref{eq:reservoir} is performed using a fourth-order Adams-Bashforth method. For an arbitrary dynamical variable $f$, the update rule over a time step $\Delta t$ is
\begin{equation}
f(t + \Delta t) = f(t) + \Delta t \left( \frac{23}{12} \frac{\partial f(t)}{\partial t} - \frac{4}{3} \frac{\partial f(t-\Delta t)}{\partial t} + \frac{5}{12} \frac{\partial f(t-2\Delta t)}{\partial t} \right).
\label{eq:adams}
\end{equation}


\begin{table}[h]
\centering
\caption{System parameters used in the numerical simulations.}
\label{tab:parameters}
\begin{tabular}{lcc}
\hline
\textbf{Parameter} & \textbf{Symbol} & \textbf{Value} \\
\hline
Polariton effective mass & $m_{\text{pol}}$ & $4.55 \times 10^{-32}$\,kg \\
Polariton-polariton interaction strength & $\alpha_1$ & $1.152 \times 10^{-20}$\,eV$\cdot$m$^2$ \\
Polariton-reservoir interaction strength & $\alpha_{1R}$ & $4.608 \times 10^{-20}$\,eV$\cdot$m$^2$ \\
Cavity photon lifetime & $\tau_{\text{cav}}$ & $1.0 \times 10^{-11}$\,s \\
Exciton reservoir lifetime & $\tau_R$ & $2.0 \times 10^{-11}$\,s \\
Stimulated scattering rate & $R$ & $2.625 \times 10^{-18}$\,eV$\cdot$m$^2$ \\
Reservoir gain coefficient & $\eta$ & $3.92 \times 10^{-11}$\,s \\
Noise amplitude & $\sigma$ & $1.18 \times 10^{-24}$\,J \\
\hline
Simulation time step & $\Delta t$ & $4.0 \times 10^{-15}$\,s \\
Spatial grid spacing & $h$ & $1.0 \times 10^{-6}$\,m \\
Number of grid points & $N \times M$ & $256 \times 256$ \\
\hline
\end{tabular}
\end{table}

The theoretical pump threshold density $P_{\text{th}}$ for a homogeneous system is defined as 
\begin{equation}
P_{\text{th}} = \frac{1}{\tau_{\text{cav}} \tau_R R},
\label{eq:pth}
\end{equation}
which follows directly from the steady-state condition of Eqs.~\eqref{eq:polariton} and \eqref{eq:reservoir} in the absence of spatial inhomogeneities and noise.

\clearpage

\section{Winding configurations $-2\pi, 0, +2\pi$ within GPE approach}

\begin{figure}[h]
	\centerline{\includegraphics[width=\columnwidth]{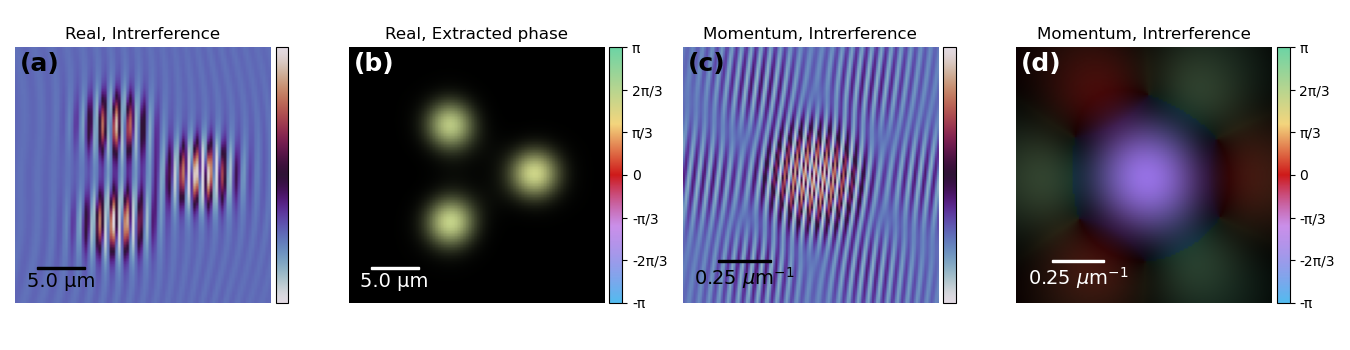}}
	\centerline{\includegraphics[width=\columnwidth]{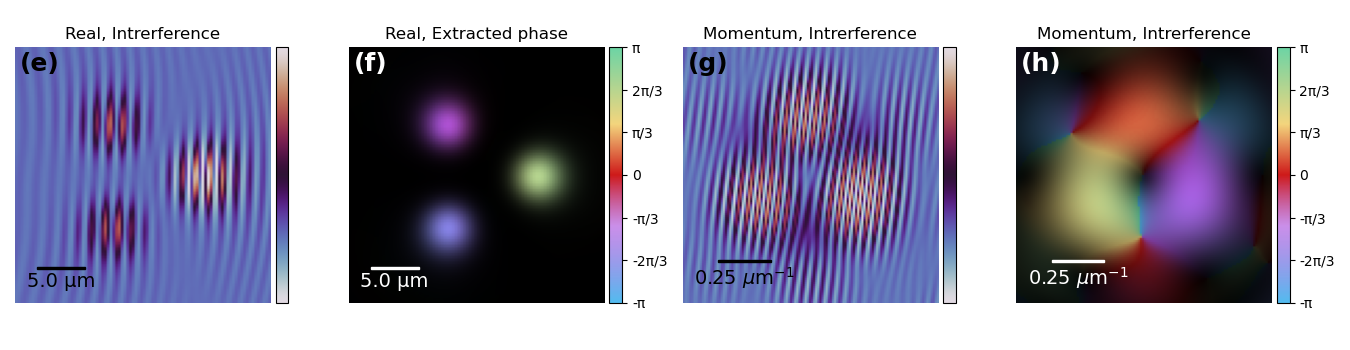}}
	\centerline{\includegraphics[width=\columnwidth]{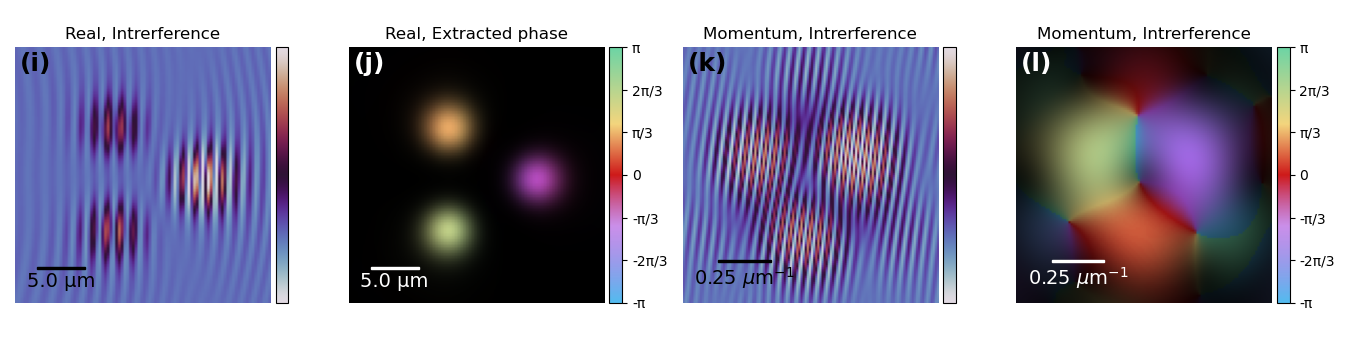}}
	\caption{Results of Gross-Pitaevskii equation simulations for observation of distinct winging sign $l = -1, 0 , +1$. Real and momentum space interferogram and retrieved phase for polariton triad are shown in real and momentum space. Row (a)-(d) if for the all-in-phase configuration at high pump power. The rows of panels (e)-(h) and (i)-(l) are for opposite winding directions of global phase in the frustrated configuration at low pump power.}
	\label{fig:S5}
\end{figure}

\clearpage

\section{OAM decomposition}

\begin{figure}[h]
    \includegraphics[width=0.32\columnwidth]{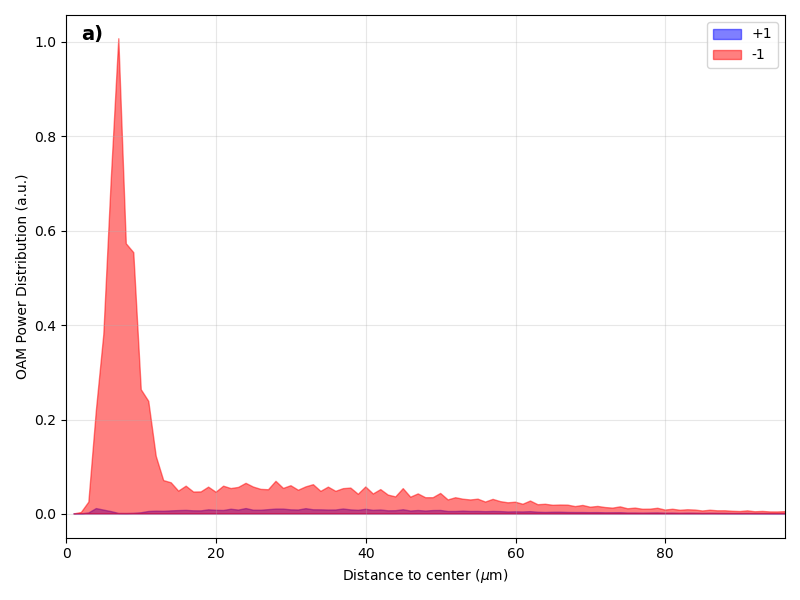}    \includegraphics[width=0.32\columnwidth]{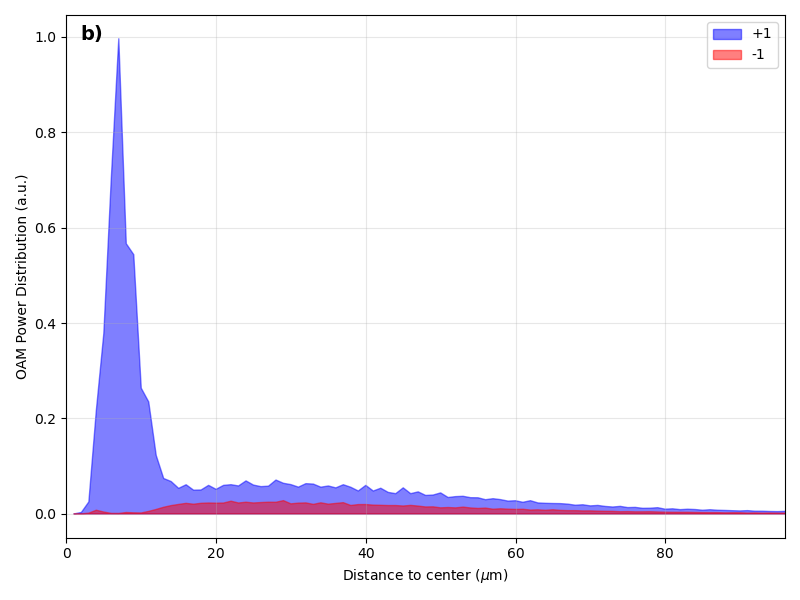}    \includegraphics[width=0.32\columnwidth]{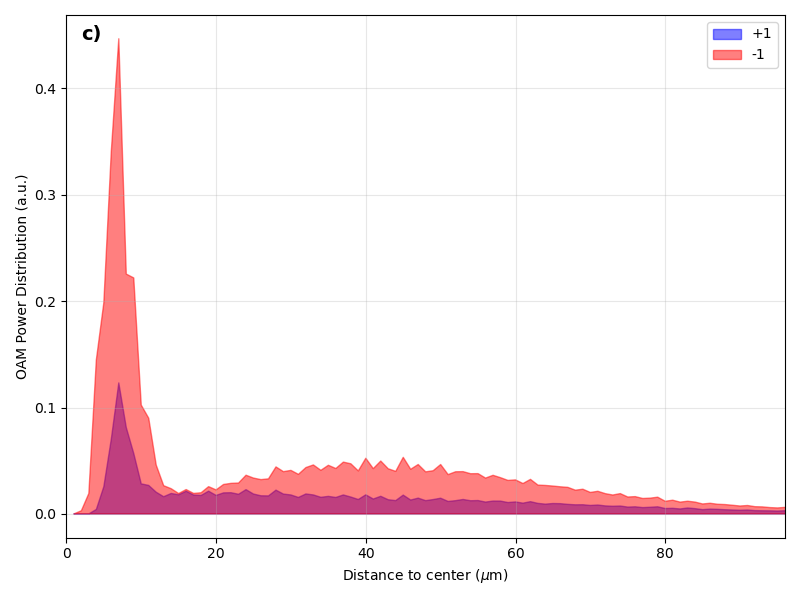}
	\caption{OAM decomposition of polariton order parameter obtained within ddGPE approach for optical potential lacking chirality. Panels a) and b) show typical situations with different -1 or +1 winding and c) is a statistically rare case with unbalanced co-existance of the two states. Difference is in random noise seed in each simulation run.}  
	\label{fig:OAMdecomp}
\end{figure}

\end{document}